**Exploring Commercial Vehicle Detouring Patterns through the Application of Probe Trajectory Data**

**Mark L. Franz, Ph.D.**
Center For Advanced Transportation Technology Laboratory (CATT Lab)
University of Maryland – College Park
College Park, MD 20742 (USA)
Email: mfranz1@umd.edu
(Corresponding Author)

**Sara Zahedian, Ph.D.**
Center For Advanced Transportation Technology Laboratory (CATT Lab)
University of Maryland – College Park
College Park, MD 20742 (USA)
Email: szahedi1@umd.edu

**Dhairya Parekh**
Center For Advanced Transportation Technology Laboratory (CATT Lab)
University of Maryland – College Park
College Park, MD 20742 (USA)
Email: dhairyap@umd.edu

**Tahsin Emtenan, Ph.D.**
Center For Advanced Transportation Technology Laboratory (CATT Lab)
University of Maryland – College Park
College Park, MD 20742 (USA)
Email: tahsin@umd.edu

**Greg Jordan**
Center For Advanced Transportation Technology Laboratory (CATT Lab)
University of Maryland – College Park
College Park, MD 20742 (USA)
Email: gjordan1@umd.edu

Keywords:

Word Count: 5,104 words + 3 table (250 words per table) = 5,854 words

*Submitted July 31, 2023*

**ABSTRACT**

Understanding motorist detouring behavior is critical for both traffic operations and planning applications. However, measuring real-world detouring behavior is challenging due to the need to track the movement of individual vehicles. Recent developments in high-resolution vehicle trajectory data have enabled transportation professionals to observe real-world detouring behaviors without the need to install and maintain hardware such as license plate reading cameras. This paper investigates the feasibility of vehicle probe trajectory data to capture commercial motor vehicle (CMV) detouring behavior under three unique case studies. Before doing so, a validation analysis was conducted to investigate the ability of CMV probe trajectory data to represent overall CMV volumes at well-calibrated count stations near virtual weigh stations (VWS) in Maryland. The validation analysis showed strong positive correlations (above 0.75) at all VWS stations. Upon validating the data, a methodology was applied to assess CMV detour behaviors associated with CMV enforcement activities, congestion avoidance, and incident induced temporary road closures. In each case study, probe trajectory data successfully captured detouring of CMVs.

## INTRODUCTION

Commercial motor vehicles (CMVs) play a vital role in the movement of goods and services within communities and across the nation, contributing significantly to economic activity. Understanding how CMVs choose routes and navigate detours is of paramount importance for optimizing transportation networks, enhancing safety, and promoting efficient freight movement.

The primary objective of this study was to leverage vehicle trajectory data to investigate the detouring behavior of CMVs under different scenarios. Three case studies were considered to demonstrate how such patterns can be revealed using trajectory data from CMVs:

1. CMV detouring close to Truck Weigh and Inspection Stations (TWIS) and the impact of CMV enforcement initiatives,
2. CMV detouring to avoid recurring congestion and the impact of intermittent holding of merging traffic to mitigate recurring congestion on US-50 in Maryland close to the Chesapeake Bay Bridge, and
3. CMV detouring due to a major crash on I-95 in Maryland between Washington, D.C. and Baltimore, Maryland.

Before investigating these cases studies, the utilization of virtual weigh station (VWS) data was employed to show how the probe trajectory data, as a sample, can represent the temporal and spatial pattern of the overall CMV traffic stream.

Detouring is a key traffic management strategy that enables temporary mobility improvements and promotes uninterrupted flow in the corridor, especially when the original roadway becomes impassable (1). Detours may also be taken by drivers for other reasons, such as avoiding roadwork, accidents, unexpected traffic congestion, and/or enforcement zones. Drivers may choose to explore alternative routes to reach their destinations faster or avoid specific areas with known traffic issues. Various unplanned event occurrences may require the consideration of detouring. For instance, detours were successfully implemented to manage the regular flow of vehicles that typically used the closed section of the Santa Monica Freeway during a three-month period following the 1994 Northridge earthquake (2). In southeast Florida, during major reconstruction on Interstate 95, detours were employed to manage peak traffic and ensure continuous flow (3). Detour plans also play a crucial role during planned special events, directing traffic away from affected areas and ensuring smoother traffic flow (4). Some drivers of CMVs use detours as a means to evade enforcement. The prevalence of evasive tendencies among certain drivers, such as traveling during hours when automatic weigh stations are closed, or even taking detours to avoid these stations completely, have been noted in previous studies (5,6,7). Detouring can have a significant impact on traffic safety. Khattak et al. conducted a study using a unique dataset to examine the relationship between work zone attributes, severe injuries, and total harm in crashes. The study found that crashes occurring in work zones where the roadway is closed, necessitating a detour on the opposite side of the median, are associated with a 38.5% increase in the likelihood of injuries (8).

CMVs play a crucial role in transportation, but also pose unique challenges to traffic safety. In recent years, there has been a notable rise in large-truck crashes, particularly those resulting in fatalities, driven by the increasing demand for freight and the growth of the Transportation Service Index (9). The study by Chen et al. found that a higher proportion of taxi and light-goods vehicles was associated with increased rates of slight-injury crashes, while a decrease in the slight injury crash rate was observed for medium- and heavy-goods vehicles (10). Several safety considerations should be taken into account when it comes to detouring for commercial motor vehicles. It is essential to ensure that the geometrics of the alternate route can reasonably accommodate detouring CMVs. This includes identifying any commercial vehicle restrictions such as height restrictions and/or limited turning radii that might hinder certain vehicles from safely using the detour. Additionally, the impact of steep upgrades or downgrades on the alternate route must be thoroughly examined, especially considering potential safety issues during adverse weather conditions (1).

Different data sources have been used in previous studies to evaluate the impact of detouring. Hainen et al. conducted a study demonstrating the practicality of utilizing data from Bluetooth technologies to evaluate operations following an unforeseen bridge closure in northwestern Indiana. The study effectively employed Bluetooth data to assess travel times on four different alternate routes (11). Bluetooth probe tracking was employed in another study to evaluate driver diversion rates from a rural interstate highway work zone located along I-65 in northwestern Indiana (12). Agencies in Indiana have also utilized segment-based probe-vehicle data to monitor queuing and congestion on a detour route that lacks conventional intelligent transportation system infrastructure. This approach was implemented during an unanticipated closure on Interstate 65 (13). In a recent study conducted by Jairaj et al., connected vehicle data was employed to identify the primary alternate route choices and affected interstate exits during an ongoing or historical incident (14). The identification of vehicle detours using the mentioned data sources requires extensive analysis and considerable labor. To overcome this, Kawasaki et al. used machine learning techniques to detect the detours of commercial vehicles in western Japan during heavy rainfall (15).

Previous research has primarily focused on examining the overall impact of detouring in specific scenarios using various data sources. However, this study takes a different approach by utilizing the same dataset to explore detouring behaviors across more generalized scenarios. To achieve this goal, three compelling case studies are presented in this paper, highlighting how trajectory data can be used to reveal CMV detouring patterns in various situations. In each case study, INRIX CMV data was analyzed using the University of Maryland's Trip Analytics platform (16). In the first case study, CMV detouring behaviors concerning TWISs and the influence of enforcement initiatives are examined. TWISs are critical locations where CMVs are subject to weight and safety inspections, often leading to potential delays and detours. The patterns of CMV detours near TWISs and the impact of enforcement actions on their routing decisions were uncovered through the analysis of probe trajectory data. The second case study focuses on CMV detouring to avoid recurrent congestion, with a particular emphasis on the effects of the intermittent holding of merging traffic near US-50, close to the Chesapeake Bay Bridge in Maryland. The analysis of probe trajectory data in this case study identified the detouring patterns of CMVs in response to congestion and explored the effectiveness of intermittent holding of merging traffic as a congestion mitigation strategy. In the third case study, CMV detouring in the event of a major crash on I-95 in Maryland was investigated. Incidents, such as crashes or road closures, can significantly impact traffic flow and may force CMVs to seek alternative routes on the fly. Insights into the detouring behavior of CMVs during major incidents and the assessment of the efficiency of diversion strategies were gained through the analysis of probe trajectory data in such scenarios.

**DATA**
**Data Description**
The two main data sources employed in this study were CMV probe trajectory data provided by INRIX and VWS data from the Maryland Department of Transportation (MDOT). The following section offers an overview of this data, detailing its characteristics and how it was collected and processed for analysis.

*Probe Trajectory Data*
Probe data has predominantly served as a valuable source for obtaining speed and travel time reliability information (17). Over time, advancements in technology have enabled the collection of more comprehensive and detailed data, allowing probe data to be increasingly utilized as trajectory data in recent years (18). Probe trajectory data, often referred to as connected vehicle data, has become a valuable asset for researchers and practitioners in the transportation field due to its ability to capture detailed and continuous information about the movement of vehicles. This data is typically obtained from Global Positioning System (GPS) devices installed in vehicles or mobile devices, enabling continuous tracking of vehicle movements (19). It is worth noting that only a portion of the vehicles in the general traffic stream are being used as probes for data providers at any one time. Thus, a validation analysis (details in the Data Validation section) was conducted to ensure the CMV probe trajectory data used in this study accurately represented the full CMV traffic stream.

The three case studies presented in this paper made use of INRIX freight trajectory data in Maryland for the year 2022. INRIX probe data has been widely employed in numerous transportation studies and applications (20, 21, 22). Each trajectory in this dataset is uniquely identified by a trip id associated with a CMV trip that recorded at least one GPS waypoint location within Maryland during 2022. For geospatial context, these GPS locations were conflated to OpenStreetMap (OSM) segments. Routes with missing segments or other anomalies were folded into other routes sets with essentially identical pathways (but were without the gaps or anomalies). By applying this algorithm, the continuity and reliability of the INRIX trajectory data was enhanced, thereby facilitating more accurate and comprehensive analysis in this study.

*Virtual Weight Stations Data*
VWS is an enforcement structure that allows continuous, remote monitoring of CMV at a defined location. VWS utilizes Weigh-in-Motion (WIM) technology to support enforcement efforts without impacting freight travel time. Targeted enforcement using this technology improves commercial vehicle safety while reducing costs and improving efficiency for enforcement activity. **Figure 1** depicts a VWS layout.

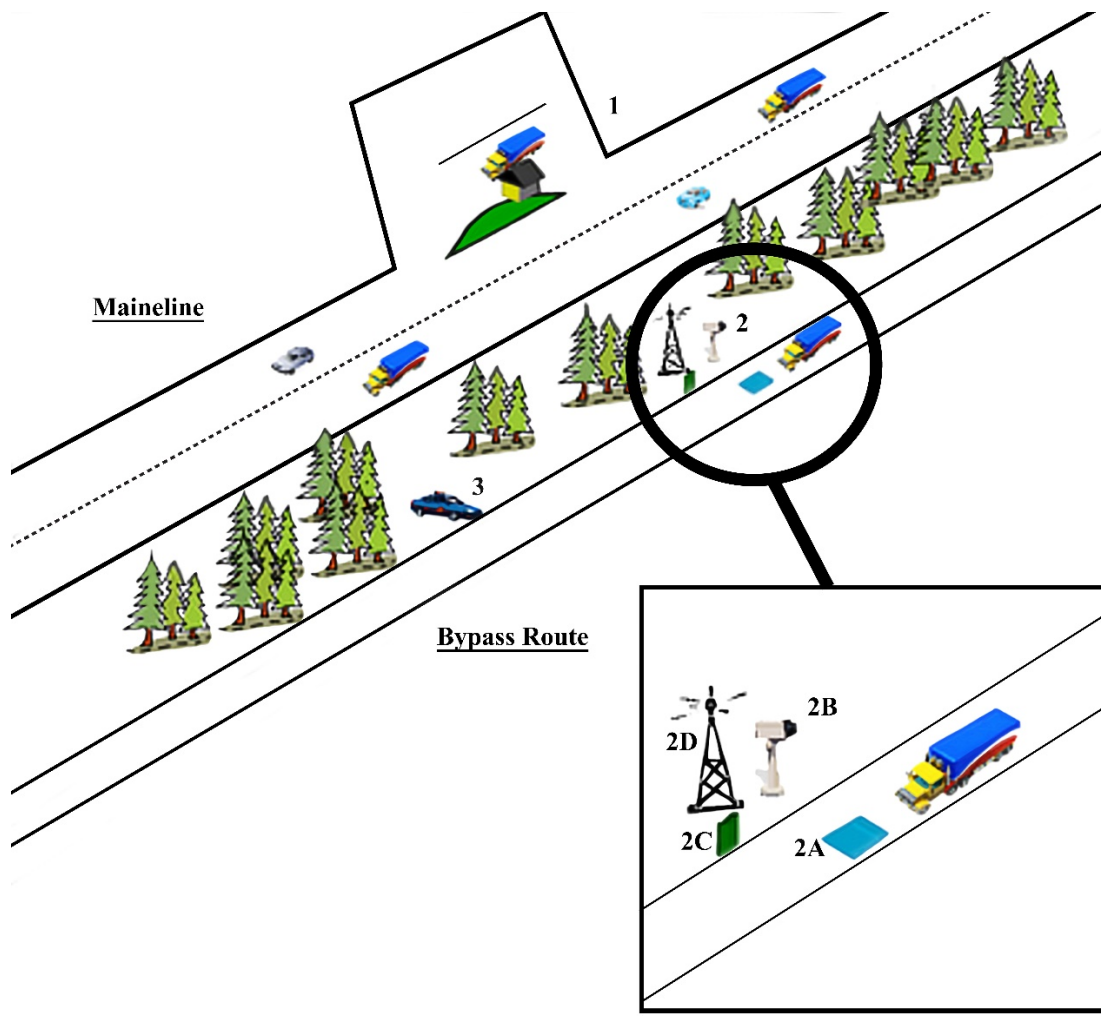

**Figure 1: VWS Physical Layout (1: Fixed Weigh Station on Primary Route, 2: VWS Deployed on a known Alternate Route, 2A: WIM Scales, 2B: Camera System, 2C: Screening Software, 2D: Communication System, 3: Mobile Enforcement Unit deployed Downstream from VWS) (23)**

When a vehicle passes over the WIM scale or sensor, it captures information about its weight, load balance over axles, speed, direction of travel, and other relevant characteristics, along with an image from the installed camera system. This data is promptly transmitted to the mobile enforcement unit for immediate action. The recorded information is archived in compliance with the state's operational security and privacy guidelines.

As of June 2023, Maryland had 26 VWS installations, with four currently non-operational, and three solely focusing on over-height violation checks. This data included attributes such as anonymized vehicle identification numbers, time of the day, vehicle class, gross weight (lbs.), speed (mph), and violations (overweight axle/tandems/bridge, length, unbalanced load, height, direction of travel, following too closely). For this study, VWS data for the non-passenger vehicle classes at each operational VWS station during 2022 was used to obtain ground truth CMV counts. **Figure 2** illustrates the locations of these active VWS stations while **TABLE 1** provides a brief description of each VWS.

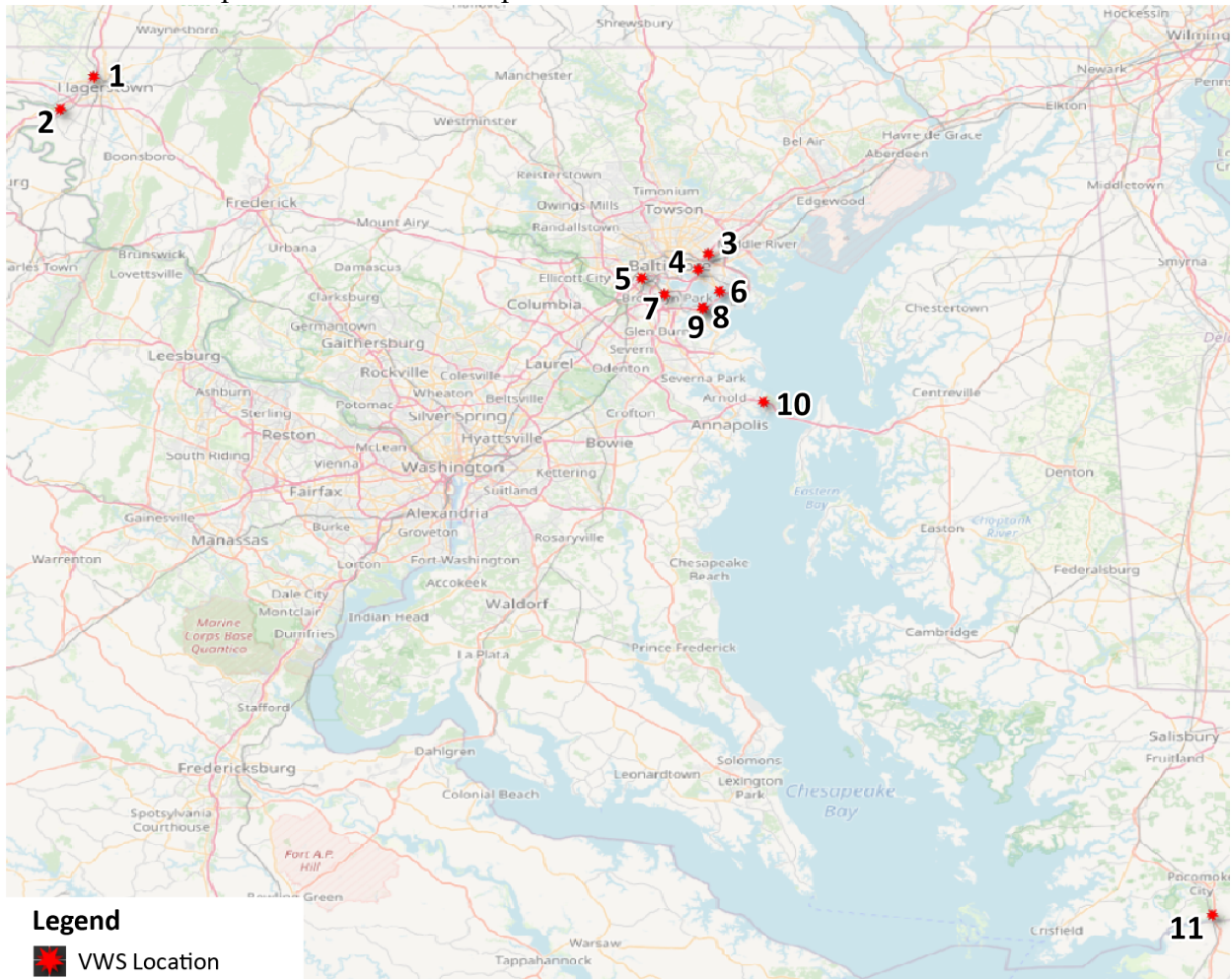

**Figure 2: Location of Virtual Weigh Stations in Maryland.**

**TABLE 1: Figure 2 VWS Locations Map Legend**

| Station Number | Station Name |
|:---:|:---|
| 1 | I-81 S at Milepost 7.6 Washington County |
| 2 | I-81 N at Milepost 1.8 Washington County |
| 3 | I-95 S at FMT Milepost 60.4 Baltimore City |
| 4 | I-895 S at Fait Ave Harbor Tunnel |
| 5 | I-95 N at Caton Ave Baltimore City |

| 6 | MD-695 S Broening Hwy Baltimore City |
| 7 | I-895 N at Milepost 6.3 Harbor Tunnel |
| 8 | I-695 N at Dock Road Baltimore City |
| 9 | I-695 S at Bear Creek Baltimore County |
| 10 | US-50 E at Whitehall Road Anne Arundel County |
| 11 | US-13 N at Tulls Corner Worcester County |

**Data Validation**

Before analyzing the case studies, this study aimed to investigate whether the probe trajectory data used to explore CMV detour patterns under different scenarios accurately represents the actual CMV traffic. To achieve this, a comparison was made between the probe trajectory data and VWS data.

Initially, unique probe trip ids at the locations of VWS stations were extracted and aggregated at the daily level. VWS data for the non-passenger class was also aggregated to obtain CMV daily counts at each active VWS station. By doing so for each day, the observed CMV counts, and probe vehicle counts at VWS locations were obtained and compared. **Figure 3** provides a sample of the daily pattern of CMV counts versus probe vehicle counts at a VWS location on SB I-95 at FMT milepost 60.4 in Maryland (Station #3 in **Figure 2**). This comparison assessed the representativeness of the probe trajectory data and its suitability for analyzing CMV detour patterns. Based on **Figure 3**, the daily pattern of probe vehicle counts closely followed the actual CMV daily counts. The drop in CMV traffic during the weekend was also clearly reflected in the probe vehicle counts.

To better quantify the correlation between probe vehicle counts and CMV counts at VWS, the study utilized the Pearson correlation coefficient. The Pearson correlation coefficient, often denoted as "r," measures the strength and direction of a linear relationship between two variables. It ranges from -1 to +1, where:

- $r = +1$ indicates a perfect positive correlation, meaning that as one variable increases, the other also increases proportionally.
- $r = -1$ indicates a perfect negative correlation, implying that as one variable increases, the other decreases proportionally.
- $r = 0$ suggests no linear correlation between the two variables.

In this context, the Pearson correlation coefficient helped determine how well the probe trajectory data aligned with the observed CMV counts at VWS locations. **Figure 4** displays the distribution of Pearson correlation coefficients for all the active VWS stations. Each box plot in this figure consists of 52 points, representing the correlation between daily probe vehicle counts and observed CMV counts over a week. According to **Figure 4**, the correlations between probe vehicle counts and observed CMV counts are consistently high and positive, with almost all values above 0.8. This indicated a strong positive correlation, suggesting that the probe trajectory data closely aligns with the actual CMV traffic patterns. The high positive correlations observed in this analysis provided evidence that CMV probe trajectory data effectively represents CMV patterns across different locations and consistently during various times. This level of correlation confirmed the data's feasibility in reflecting the overall CMV traffic patterns and established credibility for conducting detouring analysis.

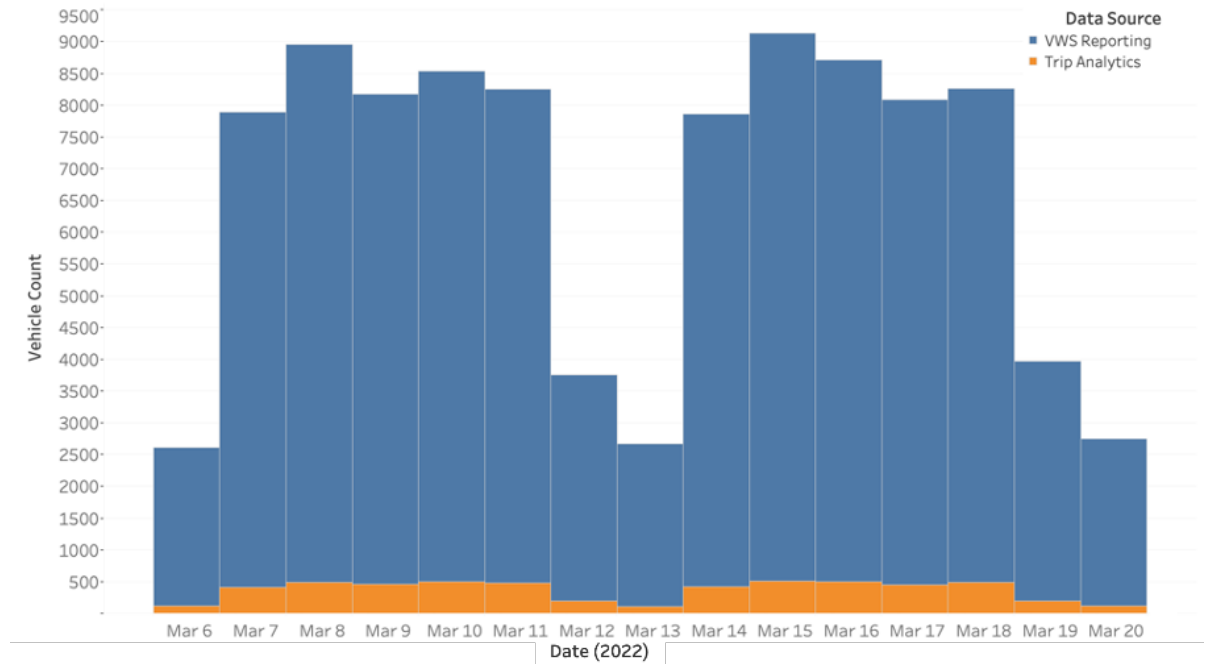

**Figure 3: Comparison of Daily CMV Counts and Probe Vehicle Counts at VWS Location on SB I-95 at FMT milepost 60.4 in Maryland.**

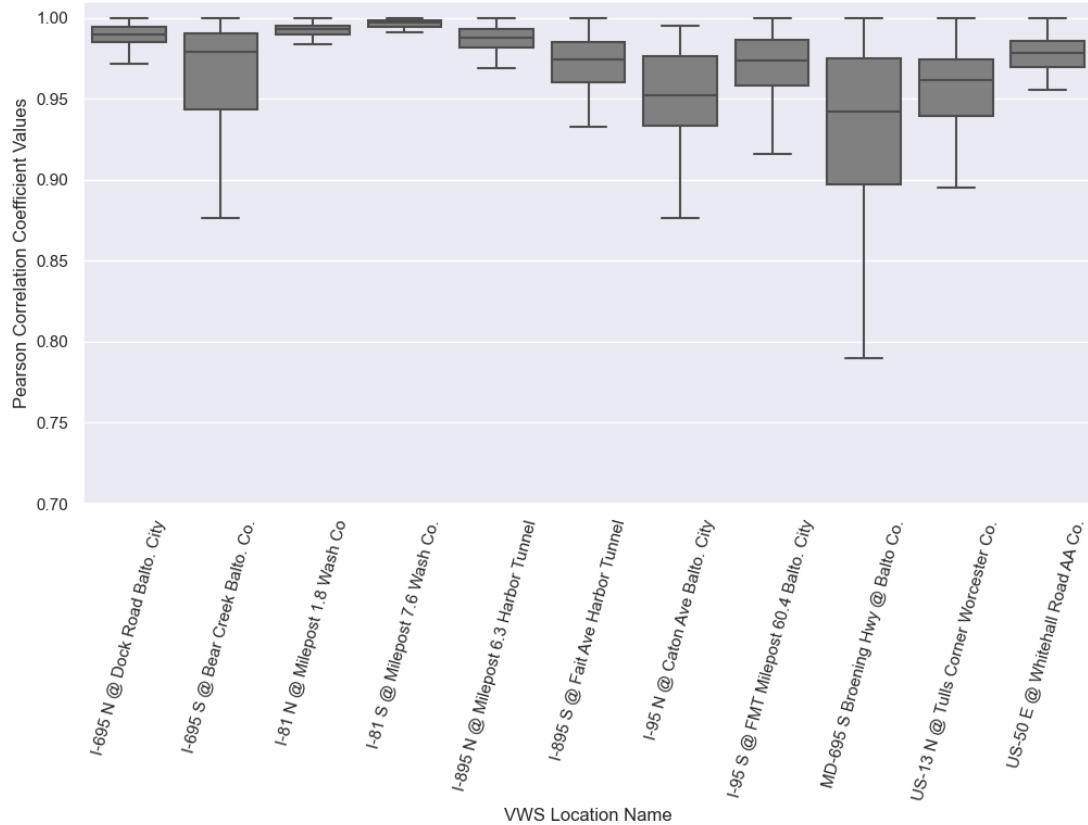

**Figure 4: Pearson Correlation Coefficients between CMV Probe Trajectory Data and Observed CMV Counts at Active VWS Stations.**

## RESULTS AND DISCSUSSION: CASE STUDIES

### Case Study 1: CMV Enforcement Initiatives

This case study focused on TWIS enforcement and the issue of detouring to avoid CMV enforcement. Specifically, this case study examined an enforcement initiative conducted by the Maryland State Police (MSP) in coordination with the Maryland Department of Transportation State Highway Administration (MDOT SHA). The initiative involved placing Truck Weight Restriction signs on Barnesville Road in Montgomery County, MD, as part of an enforcement blitz targeting CMVs bypassing the Hyattstown TWIS on I-270 (see **Figure 5**).

The objective of this one-day enforcement effort, conducted on April 11, 2022, was to address resident complaints about CMVs using Barnesville Road to avoid the Weigh Station. The case study was limited to CMVs traveling in the southbound direction between Frederick and Gaithersburg, Maryland along I-270, considering that the Southbound TWIS experiences higher traffic volume. TWIS enforcement initiatives like this aim to tackle the problem of CMVs circumventing TWISs, which can impact traffic flow and raise concerns about road safety and compliance.

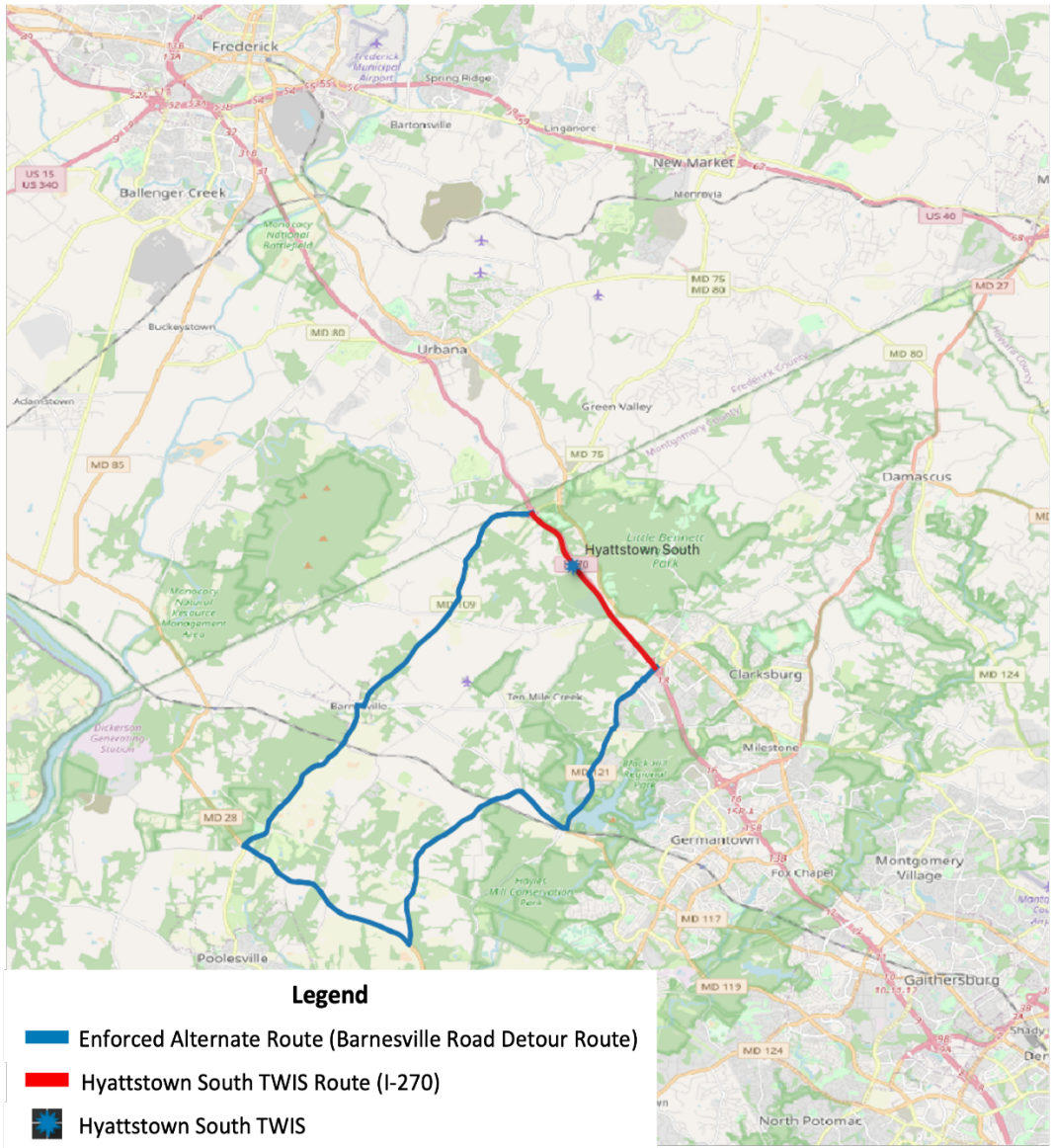

**Figure 5: Enforcement Initiative Area.**

This study utilized CMV probe trajectory data, analyzed through the Trip Analytics tool (16) to explore the detouring patterns of CMVs during the TWIS enforcement initiative. The Trip Analytics tool offered a versatile set of functionalities, enabling the extraction of specific trips passing through a narrow study area around the TWIS. Moreover, the tool provided the capability to filter and extract trips based on their origin and destination while allowing them to define specific gates for a customized analysis. By utilizing this functionality, the tool effectively filtered the probe vehicles that passed through these user-defined gates.

For this analysis, the study focused on trips that passed through a study area around the Hyattstown, Maryland TWIS on I-270 and travel from a gate located upstream of the station, near the city of Frederick, Maryland, to a gate located downstream of the weigh station, near the city of Gaithersburg, MD. **Figure 6** illustrates the study area and the locations of these gates. By utilizing the filtering capabilities of the Trip Analytics Tool, the study considers trips that could potentially use I-270 as their optimal route. This approach allows for a focused examination of CMV detouring behaviors in relation to the TWIS enforcement, providing valuable insights into the effectiveness of detouring mitigation strategies and overall traffic patterns during the enforcement period.

After applying all the filters, a total of 585 trips were extracted for analysis. The routes of these trips within the study area are presented in **Table 2**. According to the table, a majority of trips (approximately 94%) were using I-270 without passing through the SB Hyattstown TWIS. Upon inspection, the data suggested that the SB Hyattstown TWIS was open for only a few hours on this date (hence, the large portion of CMVs not stopping at the TWIS). Out of the remaining 33 trips, 21 were headed towards the Hyattstown TWIS, while the remaining 12 took alternative routes. Further analysis of these 33 trips revealed an interesting pattern. These trips occurred during specific hours, namely 8-9 AM and 3-4 PM, which coincide with the active hours of the TWIS on the date of enforcement. It is noteworthy that during these active hours, a significant number of trips (12 out of 33, i.e., 36%) avoided I-270 near the TWIS. **Figure 6** illustrates the various detouring routes that CMV drivers took that bypassed the TWIS. While one of the possible detouring routes was being enforced, detouring was still observed on secondary alternative routes.

These findings highlight the complexity of CMV detouring patterns and suggest that effective enforcement strategies should consider multiple possible detour routes used to avoid enforcement. Additionally, this case study underscores the significance of enforcement efforts during specific active enforcement hours when detouring rates appeared to be notably higher.

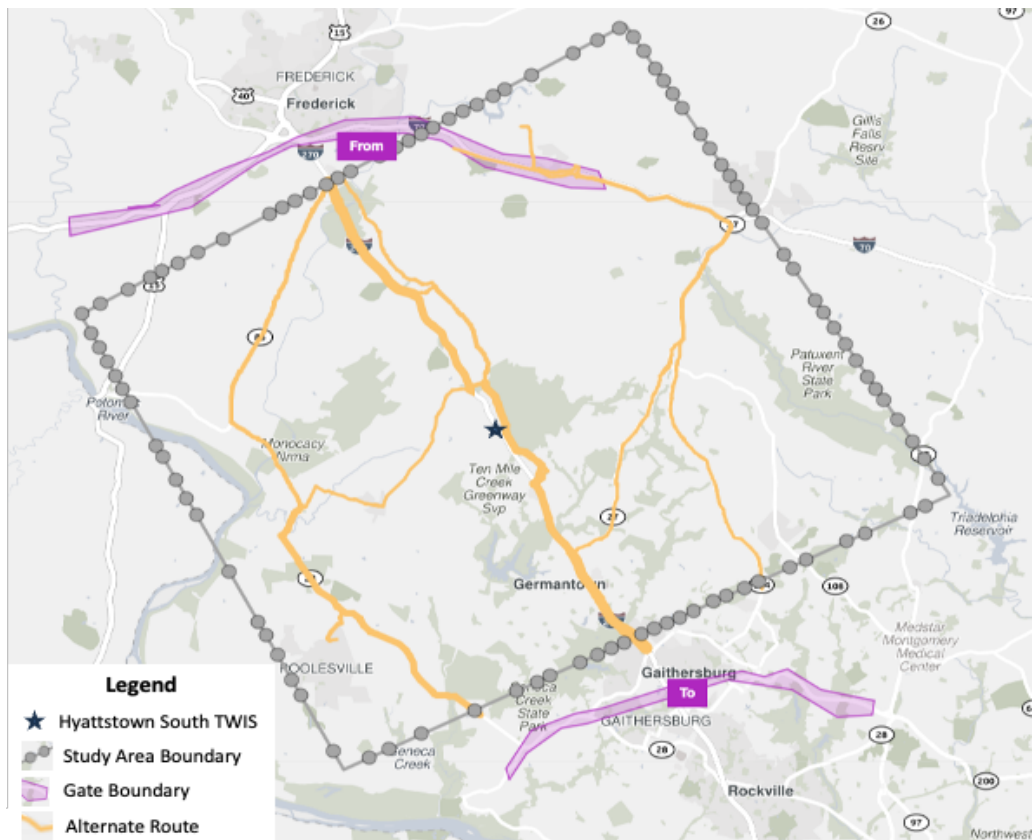

**Figure 6: Trip Analytics View, Case Study 1: Study Area, Gates Locations, and CMV Detouring Routes Likely Bypassing the Hyattstown TWIS.**

**TABLE 2 Measurement Conversion**

| Route | Number of Trips | % Trips |
|---|---|---|
| Eisenhower Memorial Highway, I-270 | 552 | 94% |
| Hyattstown South TWIS | 21 | 4% |
| Ridge Road, MD-27 | 5 | 0.9% |
| Dickerson Road, MD-28 | 3 | 0.5% |
| Frederick Road, MD-355 | 3 | 0.5% |
| Old Hundred Road, MD-109 | 1 | 0.2% |
| **Total** | **585** | **100%** |

**Case Study 2: US-50 Temporary On-Ramp Traffic Control**

The Maryland Department of Transportation State Highway Administration (MDOT SHA) implemented a temporary traffic management pilot on EB US-50 to alleviate congestion between MD-2 and the Chesapeake Bay Bridge. US-50 and the Bay Bridge serves as a critical link between the greater Washington, D.C. region and the eastern shore of Maryland. During the summer, this segment of EB US-50 is regularly congested on Thursday through Sunday due to the heavy demand to reach the beaches of Maryland's eastern shore. During these periods of congestion, drivers were frequently using two parallel arterial routes, Skidmore Drive and College Parkway, to avoid congestion on EB US-50 before the Bay Bridge (**Figure 7**). This behavior caused mobility and safety concerns to be raised by local road users.

The pilot involved intermittent holding of merging traffic at an on-ramp for EB US-50. Specifically, a temporary signal device was placed at the end of the ramp from Oceanic Drive to EB US-50 (**Figure 7**). This on-ramp serves both College Parkway and Skidmore Drive traffic and is the final on-ramp before entering the Bay Bridge.

By holding access to US-50 from these arterial roads, MDOT SHA aimed to regulate the flow of vehicles and potentially mitigate traffic congestion during the busy summer season. **Figure 7** illustrates the location of these routes and the temporary signal. The pilot's operational hours were as follows:

- Thursday, August 4th, from 12 pm to 8 pm;
- Friday, August 5th, from 12 pm to 8 pm;
- Saturday, August 6th, from 10 am to 5 pm; and
- Sunday, August 7th, from 8 am to 5 pm.

In this study, probe trajectory data was utilized to investigate how the implementation of the intermittent holding of merging traffic impacted CMV trips. While passenger trips were desired, such data was not available for this study. Once again, the Trip Analytics tool was employed to define the study area and the gates (see **Figure 8**) for the analysis. The CMV trip data for Saturday, July 30, 2022 (a week before the pilot's intermittent holding of merging traffic) and Saturday, August 06, during the pilot study from 1-3 PM, were extracted for baseline comparison. **Table 3** presents detailed information on these extracted CMV trips.

Based on **Table 3**, the analysis revealed that without the intermittent holding of merging traffic to US-50, CMVs that selected alternative routes, i.e., detoured, experienced a significant time savings (14- and 16-minute travel times on the alternative routes compared to 42 minutes on the main route). However, during the pilot study, the travel time on the detouring route was significantly closer to the main route (22 minutes travel time on the alternative route compared to

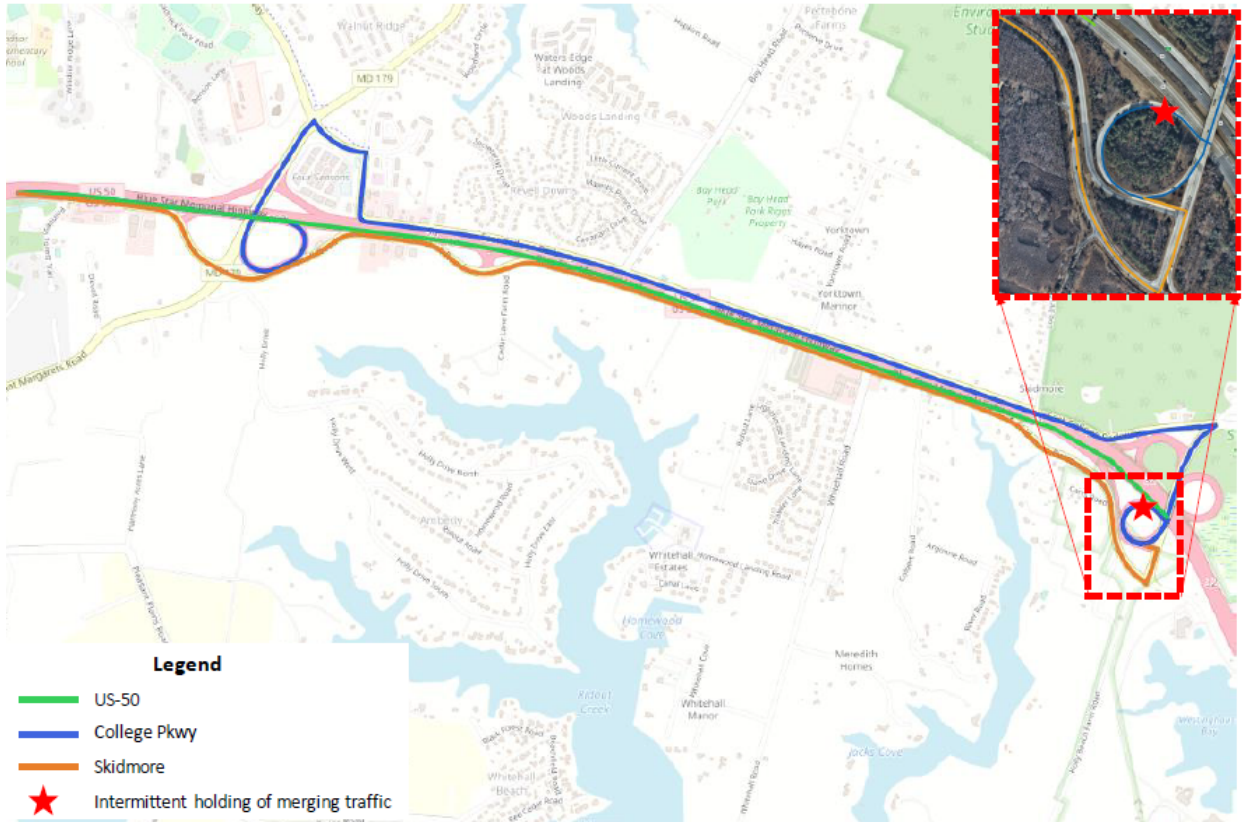

**Figure 7: US-50 Study Area.**

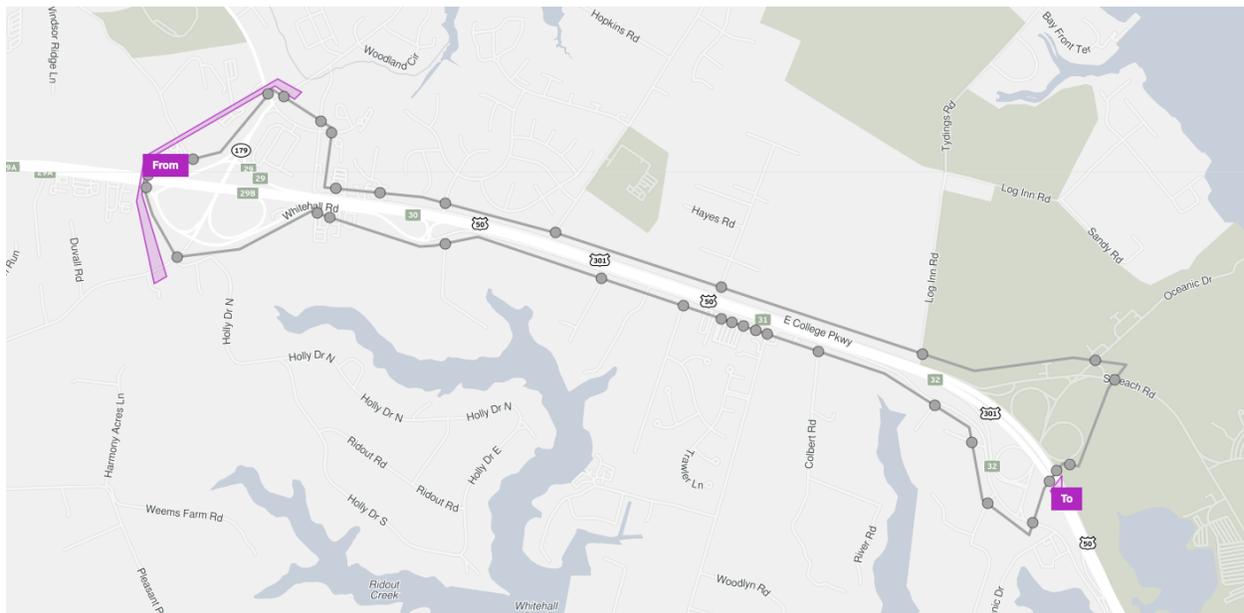

**Figure 8: Trip Analytics View, Case Study 2.**

**TABLE 3 Trip Data for Case Study 2.**

| July 30, 2022, 1-3 PM (Baseline with no traffic control) | | |
|---|---|---|
| **Route** | **Number of Trips** | **Average Travel Time (minute)** |
| EB US-50 | 6 | 42 |
| EB Skidmore Road | 2 | 16 |
| EB College Pkwy | 1 | 14 |
| **August 06, 2022, 1-3 PM (with traffic control)** | | |
| **Route** | **Number of Trips** | **Average Travel Time (minute)** |
| US-50 | 6 | 25 |
| Skidmore | 2 | 22 |
| College Pkwy | 0 | - |

26 minutes on the main route). By restricting the inflow of vehicles at the control location, the associated bottleneck was mitigated which resulted in improved travel time on EB US-50. These findings indicated the desired impact of the intermittent holding of merging traffic on CMV routing patterns. The analysis, though based on a relatively small sample size, clearly demonstrated the observed impact, showcasing another valuable application of probe trajectory data in exploring CMV routing patterns. Following this analysis MDOT SHA announced that the primary off-ramp used to access the detour routes will be permanently closed (25).

**Case Study 3: I-95 Incident**
The detouring pattern of CMVs in response to a crash can offer valuable insights into CMV adaptability and route preferences during traffic disruptions. In this case study, the detouring behavior of CMVs in response to a significant crash that resulted in the closure of all northbound lanes was examined. On Tuesday, January 19, 2021, a fatal four-vehicle crash occurred on I-95 in southwest Baltimore County. The collision began when one vehicle rear-ended another while traveling north on I-95, leading to a chain reaction involving two more vehicles. As a result of the crash, one person died at the scene, and another individual was transported to a trauma hospital for treatment. The northbound lanes of I-95 were closed for four hours following the incident (24).

I-95 is the most common route used by CMVs for travel between Washington D.C. to Baltimore. The Trip Analytics tool (16) was utilized to assess the utilization of alternate routes on the day of the crash to discover that seven different detour routes were identified for this incident. By comparing the percentages of trips on each route on the day of the crash and one week after, the detouring pattern of commercial vehicles was extracted.

An analysis similar to Case Study 1 was conducted using the Trip Analytics tool. The study focused on trips that traveled from D.C. to Baltimore, passing through a gate located near the city of Silver Spring, MD, and heading towards a gate located near the city of Towson, MD. The study area and the locations of these gates are shown in **Figure 9 and Figure 10**. After applying the appropriate filters, route information for 129 trips one week after the crash and 185 trips on the day of the crash were obtained. The analysis revealed that, one week after the crash, on a regular day, **62%** of commercial vehicles chose to use I-95 as their route from D.C. to Baltimore, while **28%** used US-50, as shown in **Figure 9**. Comparing this with the day of the crash (**Figure 10**), where only **36%** of commercial vehicles opted for I-95 that was associated with an increased proportion of trips on US-50 from **28%** to **44%**. It was evident that a larger number of vehicles diverted from the I-95 route during the immediate aftermath of the crash. On the day of the crash, a significant portion of commercial vehicles took alternative routes, likely to avoid the traffic congestion caused by the crash and the subsequent lane closures. These findings shed light on the detouring behavior of commercial vehicles in response to the crash, highlighting their adaptability and flexibility in altering routes to navigate around traffic disruptions.

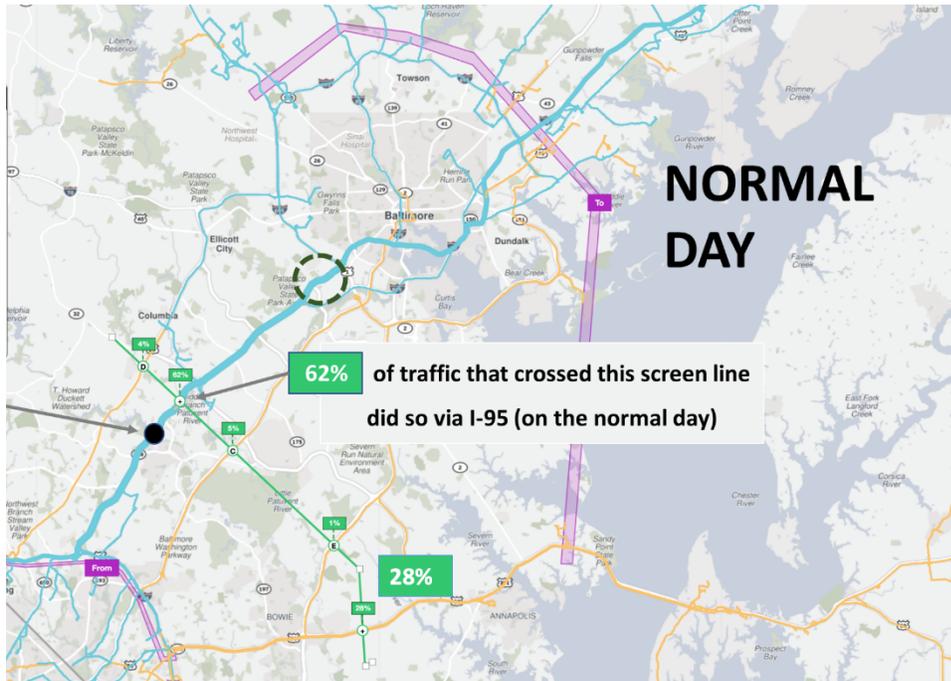

**Figure 9: Trip Analytics View, Case Study 3, One Week After the Crash**

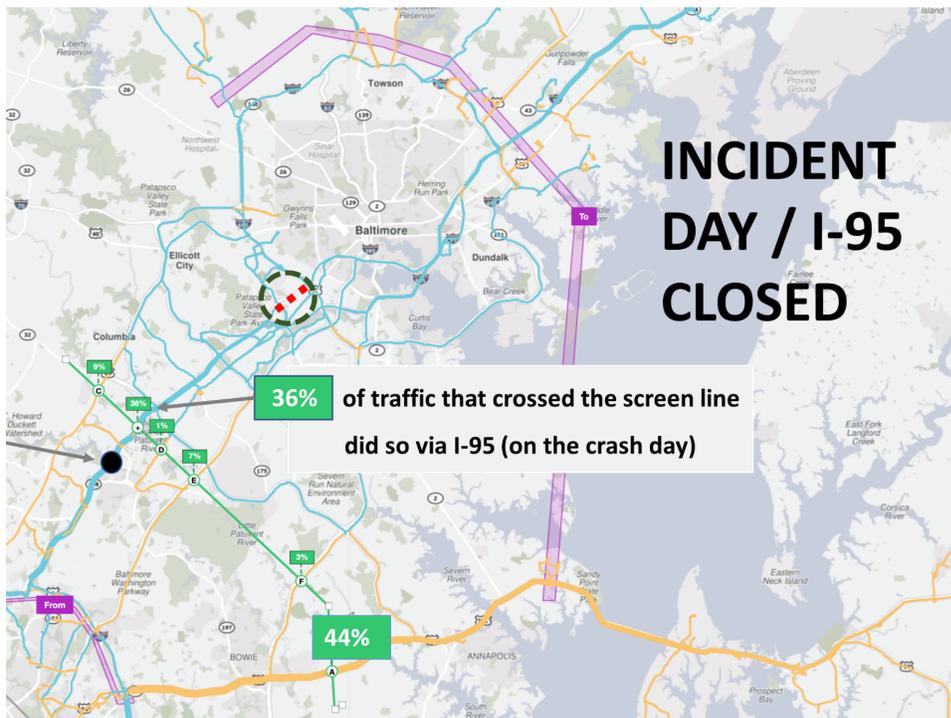

**Figure 10: Trip Analytics View, Case Study 3, on the Day of the Crash**

## CONCLUSIONS AND FUTURE WORK

This study investigated the use of probe trajectory data to assess CMV detouring behavior under three unique real-world case studies. In doing so, this research made the following contributions:

1. Summarized existing work on CMV detouring studies;
2. Found that probe trajectory data was strongly correlated with ground truth CMV demand at well calibrated VWS locations; and
3. Demonstrated that probe trajectory data can capture detouring behaviors under various traffic scenarios.

Next steps for this research include investigating additional real-world detouring scenarios, such as the I-95 bridge collapse in Philadelphia, PA. Another future path is to investigate the correlation of passenger vehicle probe trajectory data with ground truth volumes and then assess detouring behavior under several real-world case studies.

## ACKNOWLEDGMENTS

This research was partially funded by the Federal Motor Carrier Safety Administration.

The authors would like to express their sincere appreciation for the invaluable assistance provided by ChatGPT, an AI language model developed by OpenAI. ChatGPT played a significant role in paraphrasing and refining the authors' ideas and in improving clarity and coherence in the work.

## AUTHOR CONTRIBUTIONS

The authors confirm their contribution to the paper as follows:
Study conception and design: M.L. Franz and S. Zahedian
Data acquisition/collection: D. Parekh and G. Jordan
Analysis and interpretation of results: M.L. Franz, S. Zahedian, D. Parekh, T. Emtenan and G. Jordan
Draft manuscript preparation: M.L. Franz, S. Zahedian, D. Parekh, T. Emtenan and G. Jordan

# REFERENCES

1. Alternate Route Handbook. Report FHWA-HOP-06-092. FHWA, U.S. Department of Transportation, May 2006. https://ops.fhwa.dot.gov/publications/ar_handbook/index.htm. Accessed July 19, 2023.

2. Deakin AK. Potential of procedural knowledge to enhance advanced traveler information systems. Transportation research record. 1997;1573(1):35-43.

3. Edelstein R, Wolfe JA. I-95 Reconstruction: A System Maintenance of Traffic Approach. ITE Journal. 1989 Jun;59(6).

4. Suggs, E., "Festival to Test New Traffic Plan," Atlanta Journal-Constitution, April 8, 2003.

5. Cottrell BH. The avoidance of weigh stations in Virginia by overweight trucks. Virginia Transportation Research Council; 1992.

6. Cunagin W, Mickler WA, Wright C. Evasion of weight-enforcement stations by trucks. Transportation Research Record. 1997;1570(1):181-90.

7. Strathman JG, Theisen G. Weight enforcement and evasion: Oregon case study. Oregon. Dept. of Transportation. Research Group; 2002 Mar 1.

8. Khattak AJ, Targa F. Injury severity and total harm in truck-involved work zone crashes. Transportation research record. 2004;1877(1):106-16.

9. Islam M, Ozkul S. Identifying fatality risk factors for the commercial vehicle driver population. Transportation research record. 2019 Sep;2673(9):297-310.

10. Chen T, Sze NN, Chen S, Labi S, Zeng Q. Analysing the main and interaction effects of commercial vehicle mix and roadway attributes on crash rates using a Bayesian random-parameter Tobit model. Accident Analysis & Prevention. 2021 May 1;154:106089.

11. Hainen AM, Wasson JS, Hubbard SM, Remias SM, Farnsworth GD, Bullock DM. Estimating route choice and travel time reliability with field observations of Bluetooth probe vehicles. Transportation research record. 2011;2256(1):43-50.

12. Haseman RJ, Wasson JS, Bullock DM. Real-time measurement of travel time delay in work zones and evaluation metrics using bluetooth probe tracking. Transportation research record. 2010 Jan;2169(1):40-53.

13. McNamara M, Li H, Remias S, Richardson L, Cox E, Horton D, Bullock DM. Using real-time probe vehicle data to manage unplanned detour routes. Institute of Transportation Engineers. ITE Journal. 2015 Dec 1;85(12):32.

14. Desai J, Scholer B, Mathew JK, Li H, Bullock DM. Analysis of Route Choice During Planned and Unplanned Road Closures. IEEE Open Journal of Intelligent Transportation Systems. 2022 Jun 16;3:489-502.

15. Kawasaki Y, Umeda S, Kuwahara M. Detection and analysis of detours of commercial vehicles during heavy rains in western Japan using machine learning technology. Journal of JSCE. 2021;9(1):8-19.

16. Center for Advanced Transportation Technology Laboratory (CATT Lab). Trip Analytics within the Regional Integrated Transportation Information System (RITIS). https://trips.ritis.org/. Accessed July 26, 2023.

17. Federal highway Administration, Office of Operation. Freight Performance Measure Primer. 2020. https://ops.fhwa.dot.gov/publications/fhwahop16089/appc.htm. Accessed 2023 Jul 17.

18. Zhao Y, Zheng J, Wong W, Wang X, Meng Y, Liu HX. Various methods for queue length and traffic volume estimation using probe vehicle trajectories. Transportation Research Part C: Emerging Technologies. 2019 Oct 1;107:70-91.

19. Seo T, Kawasaki Y, Kusakabe T, Asakura Y. Fundamental diagram estimation by using trajectories of probe vehicles. Transportation Research Part B: Methodological. 2019 Apr 1;122:40-56.

20. Nohekhan A, Zahedian S, Haghani A. A deep learning model for off-ramp hourly traffic volume estimation. Transportation Research Record. 2021 Jul;2675(7):350-62.

21. Zahedian S, Nohekhan A, Sadabadi KF. Dynamic toll prediction using historical data on toll roads: case study of the I-66 inner beltway. Transportation Engineering. 2021 Sep 1;5:100084.

22. Dimitrijevic B, Zhong Z, Zhao L, Besenski D, Lee J. Assessing connected vehicle data coverage on new jersey roadways. In2022 IEEE 7th International Conference on Intelligent Transportation Engineering (ICITE) 2022 Nov 11 (pp. 388-393). IEEE.

23. Federal Highway Administration (FHWA). Concept of Operations for Virtual Weigh Station, https://ops.fhwa.dot.gov/publications/fhwahop09051/sec04.htm, Accessed July 19, 2023.

24. "Catonsville man dead after 4-vehicle crash on NB I-95, police say." https://www.wbaltv.com/article/luke-sounders-killed-in-crash-northbound-interstate-95-north-of-i-195-exit/35256831. Accessed July 20, 2023.

25. Eye on Annapolis. SHA to PERMANENTLY Close Whitehall Road Exit on Eastbound Route 50. June 27th, 2023. https://www.eyeonannapolis.net/2023/06/sha-to-permanently-close-whitehall-road-exit-on-eastbound-route-50/. Accessed July 26th, 2023.